\documentstyle[12pt]{article}

\begin{document}

\begin{center}
{\bf INTERPLAY BETWEEN THE ELECTRIC FIELD AND THE NONLINEAR INTERACTION\\
IN ORDERED AND DISORDERED CHAINS}\\
\vspace{1cm}

Khaled Senouci and Nouredine Zekri\footnote{Correcponding author, e-mail:
nzekri@meloo.com} \\

{\em U.S.T.O., D\'{e}partement de Physique, L.E.P.M.,\\
B.P.1505 El M'Naouar, Oran, Algeria}\bigskip .{\em \\
}\bigskip
\end{center}

\baselineskip=24pt

\centerline{\bf Abstract} \bigskip

\hspace{0.33in} A simple Kronig-Penney model is used to study the effect of
nonlinear interactions on the electronic properties of ordered and
disordered electrified chains. In the case of ordered potentials, we found that
the nonlinearity suppresses the Wannier-Stark effect caused by the electric
field.
In the case of disordered potentials, the nonlinearity gives rise to a transition
from superlocalized to weakly localized states. 

\vspace{2cm} \noindent {Keywords:} Band spectrum, non-linearity interaction,
disorder, Wannier-Stark ladder effect, superlocalization. \vspace{1cm}

\newpage

\section{Introduction}

\hspace{0.33in} It was well established two decades ago that all electronic 
 states of one dimensional (1D) disordered systems are exponentially localized 
in the absence of
external fields irrespective of the amount of disorder \cite{ATAF}. However,
recently some models of disorder introducing the correlation \cite{San,Phil}
and the nonlinearity \cite{Bourb} have been shown to exhibit extended states at
particular energies. The electric field, on the other hand has been shown to
delocalize the electronic states in 1D disordered systems where the wave
function becomes power-law decaying [5-7] while for sufficiently
large field
strenghs the eigenstates become extended \cite{Del,Senou1}. Furthermore,
it can affect the backscattering and the interferences yielding a strong
enhancement the localization ( Wannier-Stark localization) \cite{Ouas}. In
a recent paper, We found that the nonlinearity can either localize or
delocalize the electronic states depending on the strengh and the sign of
the nonlinear potential \cite{Senou2}. Physically, a repulsive nonlinear
(NL) potential represents the electron-electron interaction while an attractive 
one corresponds to the electron-phonon interaction. These
interactions are important in various systems such as quantum dots,
superlattices etc. \cite{Diez}. Therfore, the electric field and the nonlinear
potential
effects can compete and their presence together in the system may lead to the
suppression of some effects such as the Wannier-Stark localization.
This is the aim of the present letter where we examine the effect of the NL
interaction on the electronic properties of a chain of potentials in the 
presence of a constant 
electric field. Note that this effect on the resonnant transmission has been
investigated by Cota et al \cite{Cot2}. These resonnances seem to change
their structure with the NL strength. However, to the best of our knowledge
this effect on the nature of the eigenstates has not been investigated
before. \newpage

\section{Model description}

\hspace{0.35in} The model studied in this letter is defined by the
following nonlinear Schrodinger equation \cite{Cot2}

\begin{equation}
\left\{ -\frac{ d^{2} }{ dx^{2} } + \sum_{n} ( \beta_n + \alpha \left| \Psi
(x) \right| ^{2} ) \delta(x-n) -eFx\right\}\Psi (x) = E\Psi (x)
\end{equation}

\noindent Here $\Psi (x)$ is the single particle wavefunction at $x$, $\beta
_{n}$ the potential strength at the $n-th$ site, $\alpha $ the nonlinearity
strength and $E$ the single particle energy in units of $\hbar ^{2}/2m$ with 
$m$ the electronic effective mass and $F$ the electric field. The electronic
charge $e$ and the lattice parameter $a$ are taken here for simplicity 
to be unity. The two ends of the system are assumed to be connected
ohmically to ideal leads (where the electron moves freely) and maintained at
a constant potential difference $V=FL$. The potential strength $\beta _{n}$
is uniformly distributed between $0$ and $W$ in the case of potential barriers
and between $-W$ and $0$ in the case of potential wells ($W$ being the degree of
disorder). Equation (1) can be mapped by means the Poincar\'{e} map
representation in the ladder approximation (i.e, when the field can be approximated
as constant between two consecutive sites \cite{Ouas}. This approximation is
valid for $eFa\ll E$) to the following recursive equation \cite{Cot2}

\begin{equation}
\Psi_{n+1} = \left[\cos\ k_{n+1} + \frac{k_{n}\sin\ k_{n+1}}{k_{n+1}\sin\
k_{n}} \\
cos\ k_{n} +(\beta_{n}+\alpha|\psi(x)|^{2})\frac{\sin\ k_{n+1}}{k_{n+1}}%
\right]\Psi_{n}-\frac{k_{n}\sin\ k_{n+1}}{k_{n+1}\sin\ k_{n}}\Psi_{n-1}
\end{equation}

\noindent where $\Psi _{n}$ is the value of the wavefunction at site $n$ and 
$k_{n}=\sqrt{E+Fn}$ is the electron wave number at the site $n$. The
solution of equation (2) is carried out iteratively by taking the two
initial wave functions at sites $1$ and $2$ of the ideal leads : 
$\Psi _{1}=$ $\exp(-ik)$ and $\Psi _{2}=$ $\exp (-2ik)$. We consider here 
an electron with a wave number 
$k$ incident at site $N+3$ from the right side (by taking the chain length $L=N$,
i.e. $N+1$ scatterers ). The transmission coefficient ($T$) reads

\begin{equation}
T=\frac{k_{0}}{k_{L}}\frac{|1-exp(-2ik_{L})|^{2}}{|\Psi_{N+2}-\Psi_{N+3}
exp(-ik_{L})|^{2}}
\end{equation}

\noindent where $k_{0}=\sqrt{E}$ and $k_{L}=\sqrt{E+FN} $.

\section{Results and discussion}

\hspace{0.33in} In this section we examine in a first step the effect of
nonlinearity on the energy spectrum of a periodic system in the presence of
an electric field. We choose in this case $\beta =1$, $F=0.01$ and $L=500$. For 
linear systems ($\alpha =0$), the electric field seems to 
narrow the allowed bands because of the Wannier-Stark localization. Indeed, in 
this case the transmission coefficient has been shown to decrease abruptly near the
band edges 
while Bloch oscillations appear \cite{Zekri}. The nonlinearity, on the
other hand 
was found to delocalize, under certain conditions, the electronic states
in periodic systems in the
sense that the allowed bands become larger and the gaps get narrowed \cite
{Zekri2}.

\hspace{0.33in} Figure 1 shows the effect of the NL on a periodic chain of
potential barriers 
in the presence of an electric field. We in particular observe for increasing 
$\alpha <0$ , an
increase of the transmission coefficient in the regions localized by
the electric field (i.e. Wannier-Stark localization). This field induced
localization tends to diappear for a given NL strength. On the other hand,  
the amplitude of the Bloch oscillations observed in the linear case 
(solid line) seems to decrease. This delocalization is however not observed 
if we consider 
periodic potential wells whith repulsive NL, although we found recently that
this type of NL delocalizes the electronic states in the gap for such systems
\cite{Senou2}. This surprising effect may come from the unstabilities (strong drop
of the transmission) observed at certain length scales where any amount of the NL
potential strength enhances the localization \cite{Senou2}. These unstabilities 
should appear at larger length scales for the potential barriers.  

\hspace{0.33in}Let us now examine the effect of NL interactions on disordered
chains in the presence of an electric field. It was shown that the wave
function becomes power-law decaying in the presence of an electric field 
\cite{Sou,Del,Cot1}. On the other hand, the electric field was also found in
certain cases to modify
the scaling of the transmission in jumps with a behavior as 
$exp(-L^{\gamma })$ (with $\gamma >1$ and $L$ the length scale) between them
\cite{Ouas}. 
This case  was shown to correspond to a negative
differential resistance \cite{Zekri}. Figure 2 shows the transmission
coefficient versus the chain length in the case of disordered potential wells. 
We choose $E=5$,$F=0.015$ and $W=2$ with an ensemble averaging over  
2000 samples (sufficient for an accuracy about $1\%$). We observe clearly
that the superlocalization before the first jump tends to be suppressed in the
presence of a repulsive NL ($\alpha >0$) and the eigenstates become power-law
decaying. The same behavior can be observed in the case of potential
barriers (not shown here to avoid a lengthy 
paper). We note here that for almost cases the unstabilities of $T$ discussed
above \cite{Senou2} appear after the first jump of $T$. Therefore, we restricted
ourselve to the first jump. 

\hspace{0.33in}In figure 2 we observed also a characteristic length $l_{c}$
separating the superlocalized states for small lengths from the power-law decaying 
ones for larger length scales. This caracteristic length seems to decrease 
logarithmically with the NL strength in the case of disordered potential wells while
it decreases more rapidly for potential barriers (see Fig.3). \newpage 

\section{Conclusion}

\hspace{0.33in} We studied in this letter the effect of nonlinearity on
electrified periodic and disordered chains using a simple Kronig-Penney
model. We found that in periodic potential barriers, the nonlinearity 
contributes to the delocalization of the Wannier-Stark localized states induced
by the electric field. In the
case of disordered systems, we found that the superlocalization observed recently in
such systems in the presence of an electric field \cite{Ouas}
is suppressed progressively by the NL interaction and
the wave functions become power-law decaying above a characteristic
length $l_{c}$ (which seems to decrease also at least logarithmically with
nonlinearity). However, beyond a
certain length scale (corresponding after the first jump), any
amount of the NL interaction destroys the transmission in certain samples
instead of enhancing it, due to the unstability observed in nonlinear
systems. Most probably this unstability predicts very interesting statistical
properties of the transmission in such systems 
and should be carefully examined. This investigation should be the subject of a
forthcoming paper.

\newpage

\begin{center}
{\bf Figure Captions}
\end{center}

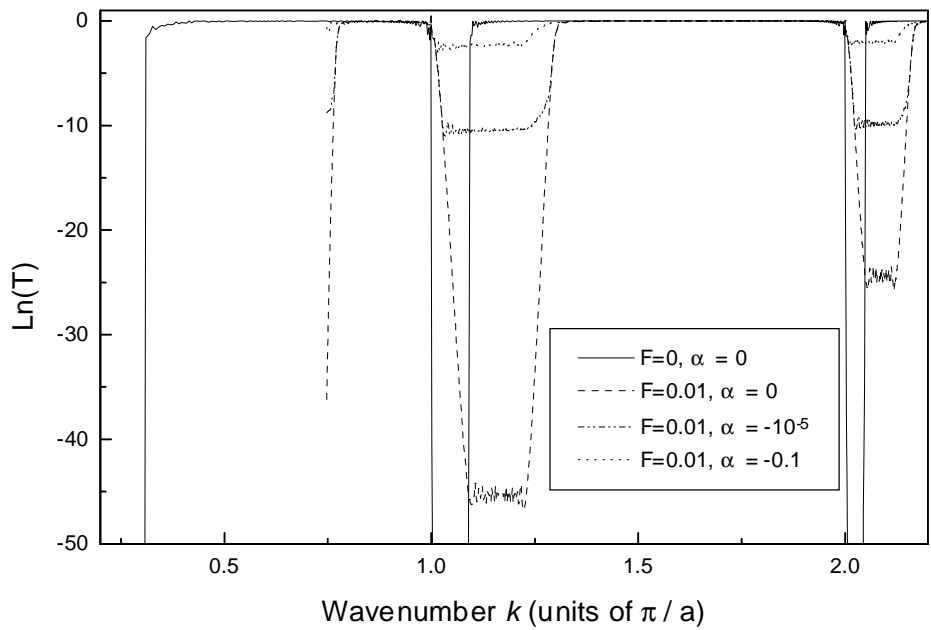
\bigskip {\bf Fig.1} $-Ln(T)$ versus the Wavenumber $k$ in units of $\pi /a$ for
ordered systems with potential barriers ($\beta =1$), a length scale $L=500$ and
$E=1$. Effect of the NL interaction.

\bigskip 
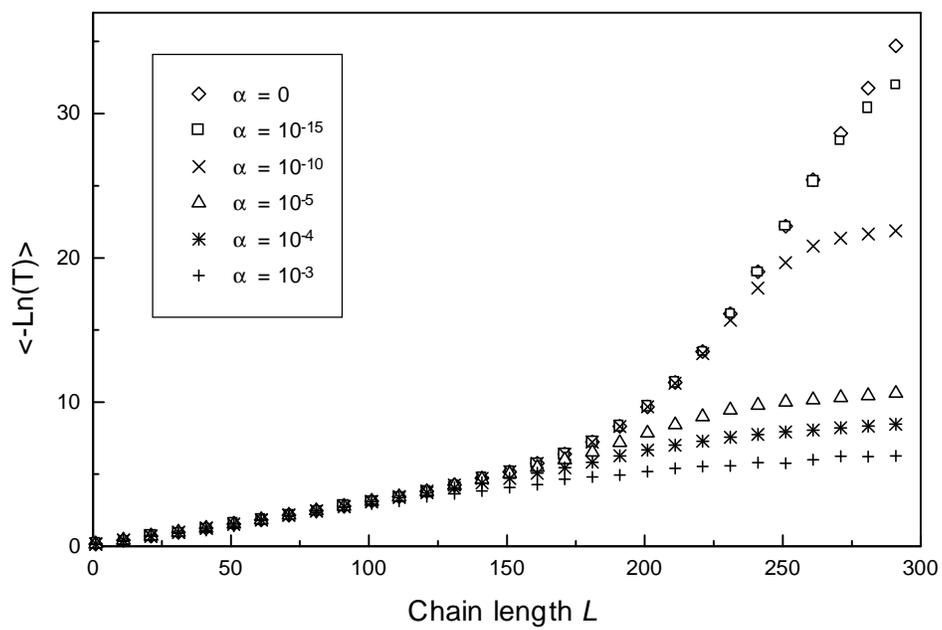
{\bf Fig.2 } $<-Ln(T)>$ versus $L$ for disordered potentials wells with
$E=5$, $W=2$, $F=0.015$. Effect of the NL interaction.

\bigskip
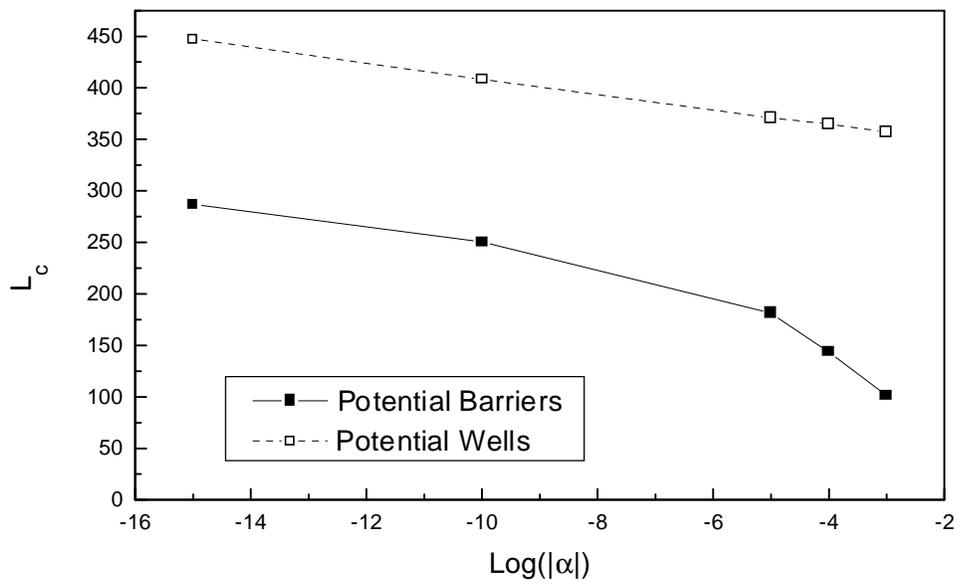
{\bf Fig3 }$L_{c}$ versus $Log(|\alpha|)$ for disordered potential barriers and
wells for the same parameters as Fig.2.

\end{document}